\documentclass[12pt,a4paper]{article}

\usepackage{epsfig}
\usepackage{amssymb}

\newcommand{\beq}[1]{
\begin{equation}\label{#1}}
\newcommand{\eeq}{\end{equation}}
\newcommand{\bea}[1]{
\begin{eqnarray}\label{#1}}
\newcommand{\eea}{\end{eqnarray}}

\date{}
{ 
 \title{\bf Scale choice \& collinear contributions to Mueller-Navelet jets at LHC energies}

\author{  F. Caporale$^1$, B. Murdaca$^1$, A. Sabio Vera$^2$, C. Salas$^2$ 
\bigskip \\ { \normalsize
 $^1$ Dipartimento di Fisica, Universit{\`a} della Calabria \&}\\ { \normalsize
Istituto Nazionale di Fisica Nucleare, Gruppo collegato di Cosenza,}
 \\ {\normalsize Parc Cient{\' \i}fic, E-46980 Paterna, Valencia, Spain.}\\ \normalsize
$^2$ Instituto de F{\' \i}sica Te{\' o}rica UAM/CSIC \& U. Aut{\' o}noma de Madrid, \\  \normalsize   
E-28049 Madrid, Spain
}

}

\begin{document}


\maketitle


We investigate the stability under variation of the renormalization, factorization and energy scales entering the calculation of the cross section, at next-to-leading order in the BFKL formalism, for the production of Mueller-Navelet jets at the Large Hadron Collider, following the experimental 
cuts on the tagged jets. To find optimal values for the scales involved in this observable it is possible to look for regions of minimal sensitivity to their variation. We show that the scales found with this logic are more natural, in the sense of being more similar to the squared transverse momenta of the tagged jets, when the BFKL kernel is improved with a resummation of collinear contributions than when the treatment is at a purely next-to-leading order. We also discuss the good perturbative convergence of the ratios of azimuthal angle correlations, which are quite insensitive to collinear resummations and well described by the original BFKL framework. 



\section*{Introduction \& theoretical set up}
The Large Hadron Collider (LHC), with a large center-of-mass energy $\sqrt{s}$,  
offers a unique opportunity to 
test our knowledge of gauge theories with a great accuracy. Besides perhaps opening a window to 
unknown new physics it also serves as a very useful tool to investigate scattering processes governed by the strong interaction, which, in fact, generate the bulk of the uncertainties and background events to new physics processes. The high energy limit of quantum chromodynamics (QCD) has been the subject of an intense debate in deep inelastic scattering (DIS) at lepton-hadron colliders since it dominates 
the growth with energy of the hadrons' structure functions. The LHC largely extends our ability to understand this type of physics since it reaches high values of $s$, motivating the need of a resummation of $\log{s}$ terms, and allows for the study of very exclusive observables, which are crucial to distinguish among different models giving very similar predictions for DIS observables 
(see, {\it e.g.},~\cite{Jung:2009eq}). 

\begin{figure}[h!]
\centering
\includegraphics[scale=0.7]{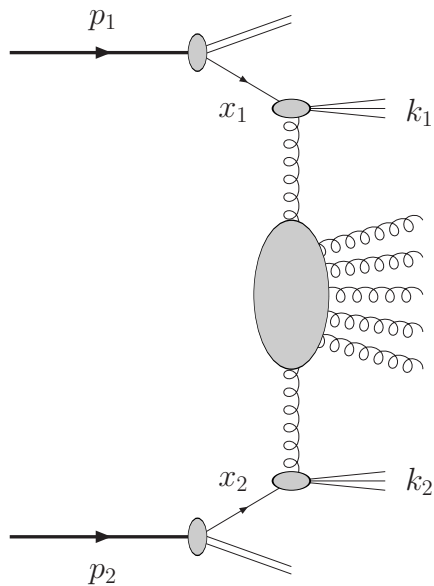}
\caption[]{
Hadroproduction of two Mueller-Navelet jets.}
\label{fig:MN}
\end{figure}
In the present work we focus on events at hadron-hadron collisions where two hard jets of similar squared transverse momentum are tagged 
with a relative rapidity $Y$ and a relative azimuthal angle $\phi$ (Mueller-Navelet jets~\cite{Mueller:1986ey}, see 
Fig.~\ref{fig:MN}). When $Y$ is large the scattering amplitudes are dominated, order by order in a perturbative expansion on the coupling 
$\alpha_s$, by terms of the form $\alpha_s Y$, which can be resummed to all orders within the 
Balitsky-Fadin-Kuraev-Lipatov formalism~\cite{BFKL1}. In principle, these Mueller-Navelet jets are interesting because 
they should manifest some sort of exponential growth with $Y$. However, this is misleading since the 
hard matrix elements are convoluted via collinear factorization with some parton distribution functions (PDFs), which damp such a behaviour. When the effect of the PDFs is so dramatic it is useful to look for 
ratios of distributions in order to remove as much as possible their contribution. Examples of such 
ratios are the average values of $\cos{(m \phi)}$, defined for integer $m$, which are functions of $Y$~\cite{DelDuca:1993mn,Stirling:1994he,Orr:1997im,Kwiecinski:2001nh}. These averages measure the azimuthal angle correlations between the two measured jets and decrease with $Y$ as $Y$ increases indicating  
that there is a large amount of soft radiation in between the two tagged jets. In principle, such excess of 
radiation cannot be easily accommodated within a fixed order calculation and experimental confirmation would be a smoking gun for the need of a resummation of the BFKL type. 

However, from a theoretical point of view, things are more complicated since any BFKL prediction is not very convergent when $m=0$~\cite{Ross:1998xw,Salam:1998tj,Vera:2005jt}. 
This corresponds to azimuthal angle averaged quantities, which are associated to the exchange of the 
so-called hard (or BFKL) Pomeron. In this case it is needed to complete the BFKL calculation by resumming, on top of the BFKL kernel, those contributions stemming from collinear regions which 
go beyond the next-to-leading order (NLO, next-to-leading approximation (NLA) or quasi-multi-Regge) kinematics~\cite{Fadin:1998py} but would be present in a putative all-orders 
calculation of the BFKL equation~\cite{Ross:1998xw,Salam:1998tj,Vera:2005jt}. Once this is done, the $m=0$ piece of any BFKL cross section is non-negative even in collinear regions (for Mueller-Navelet jets these  are defined by the $p_t^2$ 
of one of the measured jets being much higher than that of the other) and in agreement with a DGLAP analysis based on the renormalization group (RG). In the BFKL formalism, the key 
pieces which control the 
importance of these collinear contributions are the impact factors (or jet vertices in the present context), which define in which region the gluon 
Green function lies when integrating it over transverse momenta. If it is  possible to set the experimental cuts, via these impact factors, in such a way that the transverse scales during the integration do not deviate much from each other, 
then the effect of the collinear (or RG) resummation will not be very important. In general, this is not always the case and the stabilization of the BFKL series is mandatory. 

One of the targets of the present work is to show that the collinearly resummed calculation provides a 
more theoretically sound prediction than a purely NLO approach. The motivation for this reasoning is that 
when looking for a region of stability in the three-fold parameter space with renormalization $\mu_R$, factorization $\mu_F$ 
and energy scales $s_0$, we find that the NLO ``natural" scales are larger than those obtained  with a collinearly improved approach. This is a non-trivial statement since this ``naturalness" survives the influence of the 
PDFs, quite sensitive to the choice of factorization scale. A similar result was found for quite a different observable, also calculated fully at NLO: the production of two light vector mesons well separated in rapidity in hadronic collisions at high energies~\cite{mesons}. 

At this point it is important to indicate that there exist certain ratios which are quite insensitive to the collinear contributions and enjoy an excellent 
perturbative convergence within the BFKL context  (the NLO corrections are very small), these are the fractions 
\begin{eqnarray}
{\cal R}_{mn} &\equiv& \frac{\langle \cos{(m \Delta \phi)} \rangle}{\langle \cos{(n \Delta \phi)}\rangle},
\end{eqnarray}
where $\langle \cos{(m \Delta \phi)} \rangle$ are the azimuthal angle correlations defined in Eq.~(\ref{azcorr}). These ratios were proposed as the ideal BFKL observables several years ago in~\cite{Vera:2006un,Vera:2007kn} and have been shown to allow for a discrimination between BFKL and other approaches. They prove the 
conformal structure of QCD at high energies since $m$ and $n$ can be interpreted as conformal spins 
in elastic scattering, in the so-called Pomeron wave function. In this sense it is natural that they exhibit quite a different $Y$ dependence to 
that generated by more standard methods, such as Monte Carlo event generators based on angular ordering of collinear emissions, since in any other formalism there is not such a two dimensional conformal invariance. The same logic applies to any fixed order calculation. 
In~\cite{joseagusgreg} they were calculated in the $N=4$ super Yang-Mills theory (other studies of Mueller-Navelet jets can be found in~\cite{Marquet:2005vp,Marquet:2007xx}). 

Let us now introduce some of the technical details needed to produce our numerical results and to reach our conclusions on the scale dependence. The cross section for the process ${\rm Proton} \, (p_1) + {\rm Proton} \, (p_2) \to {\rm Jet}\, (k_{J_1}) +
{\rm Jet} \, (k_{J_2})+ X$, differential with respect to the
variables parameterizing the jet phase space ($dJ_i\equiv dx_{J,i}d^{D-2}k_{J,i}$) can be shown to 
 factorize in the high energy limit as a convolution of a partonic cross section with the initial proton PDFs: 
\begin{equation}
\frac{d\sigma}{dJ_1 dJ_2}
=\sum_{i,j=q,\bar q,g}\int\limits^1_0 dx_1 \int\limits^1_0 dx_2\,f_i(x_1,\mu_F)
f_j(x_2,\mu_F) \frac{d\hat \sigma_{i,j}(x_1 x_2 s,\mu_F)}{dJ_1 dJ_2}.
\label{ff}
\end{equation}
In this expression $\mu_F$ is the factorization scale and $x_{1(2)}$ are the longitudinal momentum
fractions of the initial state partons. For the hard subprocess it is convenient to write the following representation
\begin{eqnarray}
\frac{d\hat \sigma_{i,j}(x_1 x_2 s)}{dJ_1 dJ_2} &=& \nonumber\\
&&\hspace{-3.cm} \int\frac{d^2 \vec q_1}{\vec
q_1^{\,\, 2}} \frac{d\Phi_{J,1}(\vec q_1,s_0)}{dJ_1}
\int\frac{d^2 \vec q_2}{\vec q_2^{\,\,2}}\frac{d\Phi_{J,2}(\vec q_1,s_0)}{dJ_2}
\! \int\limits^{\delta +i\infty}_{\delta
-i\infty} \! \frac{d\omega}{(2\pi)^3 i}\left(\frac{x_1 x_2 s}{s_0}\right)^\omega
G_\omega (\vec q_1, \vec q_2)\, ,
\label{hard}
\end{eqnarray}
which is valid within NLO accuracy. $d\Phi_{J,1(2)}(\vec q_1,s_0)/dJ_{1 (2)}$ are the 
differential jet production vertices, calculated at NLO in~\cite{Bartels:2001ge,Bartels:2002yj} 
and more recently in~\cite{Caporale:2011cc,Ivanov:2012ms} (see also the recent derivation using Lipatov's high energy effective action in~\cite{Hentschinski:2011tz}), and $s_0$ is an energy scale to be defined and which indicates what parts of the Feynman diagrams are included in the jet vertex and which ones go to the BFKL kernel. Later in our calculation it will be convenient to make the substitution
\begin{eqnarray}
\left(\frac{x_{1} x_{2} s}{s_0}
\right)^{\omega}&=& \left(\frac{x_{J_1} x_{J_2} s}{s_0}
\right)^{\omega}\left(\frac{x_{1}}{x_{J_1}}
\right)^{\omega}\left(\frac{x_{2} }{x_{J_2}}
\right)^{\omega} \, ,
\end{eqnarray}
and to assign the last two factors in the r.h.s. to the corresponding
jet impact factors. In transverse momentum representation, defined by $\hat{\vec q} \, |\vec q_i\rangle = \vec q_i|\vec q_i\rangle$, $\langle\vec q_1|\vec q_2\rangle = \delta^{(2)}(\vec q_1 - \vec q_2)$ and 
$\langle A|B\rangle = \langle A|\vec k\rangle\langle\vec k|B\rangle = 
\int d^2k A^*(\vec k)B(\vec k)$, the equation for the gluon Green function, 
$G_\omega$, reads
\beq{Groper}
\hat 1=(\omega-\hat K)\hat G_\omega\;,
\eeq
where the kernel $\hat K$ can be written as
\beq{kern}
\hat K=\bar \alpha_s \hat K^0 + \bar \alpha_s^2 \hat K^1 + \hat K_{RG}.
\eeq
${\bar \alpha_s}$ stands for $\alpha_s N_c / \pi$, $\hat K^0$ is the leading-order (LO) kernel, $\hat K^1$ the NLO correction and $\hat K_{RG}$ contains the all-orders collinear terms which start at ${\cal O} (\alpha_s^3)$. Note that we can write it in such a simple (additive and non-transcendental) form because we have decoupled the transverse components from the longitudinal ones also for the collinearly improved kernel, how to do so was explained in~\cite{Vera:2005jt}.

The NLO solution of Eq.~(\ref{Groper}) can be written as 
\begin{eqnarray}
\hat G_\omega &=& (\omega-\bar \alpha_s\hat K^0)^{-1}+
(\omega-\bar \alpha_s\hat K^0)^{-1}\left(\bar \alpha_s^2 \hat K^1 + \hat K_{RG} \right)
(\omega-\bar \alpha_s \hat K^0)^{-1} \nonumber\\
&+& {\cal O}\left[\left(\bar \alpha_s^2 \hat K^1\right)^2\right].
\end{eqnarray}
This representation can be expanded in the basis of eigenfunctions for the LO kernel, {\it i.e.}
\begin{eqnarray}
\hat K^0 |n,\nu\rangle &=& \chi(n,\nu)|n,\nu\rangle,\\
\chi (n,\nu) &=& 2\psi(1)-\psi\left(\frac{n}{2}+\frac{1}{2}+i\nu\right)
-\psi\left(\frac{n}{2}+\frac{1}{2}-i\nu\right),
\end{eqnarray}
where
\beq{nuLLA}
\langle\vec q\, |n,\nu\rangle =\frac{1}{\pi\sqrt{2}}
\left(\vec q^{\,\, 2}\right)^{i\nu-\frac{1}{2}}e^{in\phi} \;,
\eeq
with $\phi$ being the azimuthal angle of the vector $\vec q$. In this set up, the orthonormality  condition for the eigenfunctions takes the form
\beq{ort}
\langle n',\nu^\prime | n,\nu\rangle =\int \frac{d^2 q}
{2 \pi^2 }\left(\vec q^{\,\, 2}\right)^{i\nu-i\nu^\prime -1}
e^{i(n-n')\phi}=\delta(\nu-\nu^\prime)\, \delta_{nn'}\, .
\eeq
The action of the full NLO kernel on this basis may be expressed
as follows ($\mu_R$ is the renormalization scale):
\bea{Konnu}
\hat K|n,\nu\rangle &=&
\bar \alpha_s(\mu_R) \chi(n,\nu)|n,\nu\rangle \nonumber\\
&+&\bar \alpha_s^2(\mu_R)\left(\chi^{(1)}(n,\nu)
+\frac{\beta_0}{4N_c}\chi(n,\nu)\ln(\mu^2_R)\right)|n,\nu\rangle
\nonumber \\
&+& \bar
\alpha_s^2(\mu_R)\frac{\beta_0}{4N_c}\chi(n,\nu)
\left(i\frac{\partial}{\partial \nu}\right)|n,\nu\rangle + \chi_{RG}(n,\nu)|n,\nu\rangle \;,
\eea
where the first
term represents the LO eigenvalue. The second line and first term of the third line  
stand for the diagonal and non-diagonal NLO pieces and 
\begin{eqnarray}
\hspace{-.6cm}\chi_{RG}(n,\nu) = \sum_{m=0}^{\infty} \left(\sqrt{2\left(\bar \alpha_s + a_n \bar \alpha_s^2 \right) 
+ \left( m - b_n \bar \alpha_s + \frac{1}{2} + i \nu  + \frac{ |n| }{2} \right)^2} -m - i\nu \right. && \nonumber\\
&&\hspace{-13.8cm} \left. + b_n \bar \alpha_s - \frac{1+|n|}{2} 
- \frac{\bar \alpha_s + a_n \bar \alpha_s^2}{m + \frac{1+|n|}{2} + i\nu}
- \frac{\bar \alpha_s^2 b_n}{\left(  m + \frac{1+|n|}{2} + i\nu \right)^2}
+ \frac{\bar \alpha_s^2}{\left( m + \frac{1+|n|}{2} + i\nu \right)^3}  \right)\nonumber\\
&&\hspace{-13.8cm} + \, \, \{\nu \to -\nu \},
\end{eqnarray}
is the eigenvalue of the collinear contributions with a representation in transverse momentum space in the form of a Bessel function~\cite{Vera:2005jt,Vera:2006un,Vera:2007kn}. The 
coefficients $a_n$ and $b_n$ read
 \begin{eqnarray}
a_n = \frac{4 - \pi^2 + 5\beta_0/N_c}{12} - \frac{\pi^2}{24} + \frac{\beta_0}{4N_c} \left( \psi(n+1) - \psi(1) \right) + \frac{1}{2}  \psi^\prime (n+1) \hspace{2cm} &&\nonumber\\
&& \hspace{-14cm}+ \frac{1}{8} \left( \psi^\prime \left(\frac{n+1}{2}\right) - \psi^\prime \left(\frac{n+2}{2}\right)  \right) - \frac{\delta_n^0}{36} \left(67 + 13 \frac{n_f}{N_c^3} \right) - \frac{47 \delta_n^2}{1800} \left(1 + \frac{n_f}{N_c^3} \right),
\end{eqnarray}
and
\begin{eqnarray}
-b_n = \frac{\beta_0}{8N_c} + \frac{1}{2} \left( \psi(n+1) - \psi(1) \right) + \frac{\delta_n^0}{12}  \left(11 + 2\frac{n_f}{N_c^3} \right) + \frac{\delta_n^2}{60} \left(1 + \frac{n_f}{N_c^3} \right). \hspace{1cm}
\end{eqnarray}

The function $\chi^{(1)}(n,\nu)$, calculated in~\cite{Kotikov:2000pm} (see
also~\cite{Kotikov:2000pm2}), can be presented in the form
\beq{ch11}
\chi^{(1)}(n,\nu)=-\frac{\beta_0}{8\, N_c}\left(\chi^2(n,\nu)-\frac{10}{3}
\chi(n,\nu)-i\chi^\prime(n,\nu)\right) + {\bar \chi}(n,\nu)\, ,
\eeq
with 
\begin{eqnarray}
-4 \bar \chi(n,\nu) &=& \frac{\pi^2-4}{3}\chi(n,\nu)
-6\zeta(3)-\chi^{\prime\prime}(n,\nu) +\,2\,\phi(n,\nu)+\,2\,\phi(n,-\nu) \nonumber\\
&& \hspace{-3cm} +
\frac{\pi^2\sinh(\pi\nu)}{2\,\nu\, \cosh^2(\pi\nu)} \left(
\left(3+\left(1+\frac{n_f}{N_c^3}\right)\frac{11+12\nu^2}{16(1+\nu^2)}\right)
\delta_{n0}
-\left(1+\frac{n_f}{N_c^3}\right)\frac{(1+4\nu^2)\delta_{n2}}{32(1+\nu^2)}
\right),\hspace{1cm}
\end{eqnarray}
and 
\begin{eqnarray}
\phi(n,\nu) &=& \sum_{k=0}^\infty\frac{(-1)^{k+1}}{k+(n+1)/2+i\nu}\left[\psi'(k+n+1)
-\psi'(k+1)\right.\nonumber\\
&&\hspace{-2cm}\left.+(-1)^{k+1}(\beta'(k+n+1)+\beta'(k+1)) -\frac{(\psi(k+n+1)-\psi(k+1))}{k+(n+1)/2+i\nu}
\right],
\end{eqnarray}
where $4 \beta'(z) = \psi' \left((z+1)/2 \right) -\psi' \left(z / 2\right)$ and $\mbox{Li}_2(x) =-\int\limits_0^x 
\ln{(1-t)} dt / t$.

We can now express the differential cross section for the dijet production in terms of an expansion in Fourier components in the azimuthal angle, {\it i.e.}
\beq{}
\frac{d\sigma}
{dy_{J_1}dy_{J_2}\, d|\vec k_{J_1}| \, d|\vec k_{J_2}|
d\phi_{J_1} d\phi_{J_2}}
=\frac{1}{(2\pi)^2}\left({\cal C}_0+\sum_{n=1}^\infty  2\cos (n\phi )\,
{\cal C}_n\right)\, ,
\eeq
where $\phi=\phi_{J_1}-\phi_{J_2}-\pi$, $y_{1 (2)}$ are the rapidities of the two produced jets and
\beq{Cm}
\hspace{-.6cm}{\cal C}_m = \int_0^{2\pi}d\phi_{J_1}\int_0^{2\pi}d\phi_{J_2}\,
\cos[m(\phi_{J_1}-\phi_{J_2}-\pi)] \,
\frac{d\sigma}{dy_{J_1}dy_{J_2}\, d|\vec k_{J_1}| \, d|\vec k_{J_2}|
d\phi_{J_1} d\phi_{J_2}}\;.
\eeq
The final expression reads 
\beq{C0}
\hspace{-.5cm}{\cal C}_n
= \frac{x_{J_1} x_{J_2}}{|\vec k_{J_1}| |\vec k_{J_2}|}
\int_{-\infty}^{+\infty} d\nu \, \left(\frac{x_{J_1} x_{J_2} s}{s_0}
\right)^{\bar \alpha_s(\mu_R)\chi(n,\nu)}
\eeq
\[
\hspace{-.5cm}\times \alpha_s^2(\mu_R) c_1 c_2\, \left[1
+\alpha_s(\mu_R)\left(\frac{c_1^{(1)}}{c_1}
+\frac{c_2^{(1)}}{c_2}\right)
\right.
\]
\[
\left.
\hspace{-.5cm}+\bar \alpha_s^2(\mu_R) \ln\left(\frac{x_{J_1}x_{J_2} s}{s_0}\right)
\left(\bar \chi(n,\nu)
+ \frac{\beta_0}{8C_A}\chi(n,\nu)
\left(-\chi(n,\nu) + \frac{10}{3} + \ln\frac{\mu_R^4}
{\vec k_{J_1}^2 \vec k_{J_2}^2}\right)\right)\right. 
\]
\[
\left. \hspace{-.5cm}+ \ln\left(\frac{x_{J_1}x_{J_2} s}{s_0}\right) \chi_{RG} (n,\nu)\right] \;,
\]
where 
$$c_{1 (2)} = c_{1 (2)}(n,\nu,|\vec k_{J{1 (2)}}|, x_{J_{1 (2)}},\mu_F) \hspace{.3cm}{\rm and}\hspace{.3cm} c_{1(2)}^{(1)}= c_{1 (2)}^{(1)} (n,\nu,|\vec k_{J{1 (2)}}|, x_{J_{1 (2)}},\mu_F)$$
 are, respectively, the LO and NLO contributions to the differential impact factors~\cite{Ivanov:2012ms}, projected in the $\nu$-space and convoluted with the proton PDFs. We refer the reader to~\cite{Ivanov:2012ms,Caporale:2012ih} for the explicit expressions.  We have taken the approximation of a small cone radius in the jet definition since this makes the numerical study much simpler and the final results are very similar to the equivalent ones using the exact expressions~\cite{Ducloue:2013hia,Colferai:2010wu}. 

In order to perform the numerical analysis and investigate the dependence of our results on the energy variable $s_0$, we use the representation
\beq{C0_exp}
{\cal C}^{\rm exp}_n
= \frac{x_{J_1} x_{J_2}}{|\vec k_{J_1}| |\vec k_{J_2}|}
\int_{-\infty}^{+\infty} d\nu \, \exp\biggl[(Y-Y_0)
\biggl(\bar \alpha_s(\mu_R)\chi(n,\nu)\biggr.\biggr.
\eeq
\[
\left.\left.
+\bar \alpha_s^2(\mu_R)\left(\bar \chi(n,\nu)
+ \frac{\beta_0}{8C_A}\chi(n,\nu)
\left(-\chi(n,\nu) + \frac{10}{3} \right)\right)  + \chi_{RG} (n,\nu) \right)\right]
\]
\[
\hspace{-.8cm}\times  \alpha_s^2(\mu_R)  c_1 c_2\,
\left[1 + \bar \alpha_s^2 \left( Y- Y_0 \right)  \frac{\beta_0}{8C_A}\chi(n,\nu) 
\ln\frac{\mu_R^4}{\vec k_{J_1}^2 \vec k_{J_2}^2} +\alpha_s(\mu_R)\left(\frac{c_1^{(1)}}{c_1}
+\frac{c_2^{(1)}}{c_2}\right)\right]\;,
\]
where we have introduced the rapidity variables
\begin{equation}
\label{Y0}
 Y=\ln{\left(\frac{x_{J_1} x_{J_2}}{|\vec k_{J_1}||\vec k_{J_2}|}\right)} \hspace{1cm}{\rm and}   \hspace{1cm} Y_0=\ln{\left({s_0\over |\vec k_{J_1}||\vec k_{J_2}|}\right)} \, .
\end{equation}
Note that a ``natural" value for the free scale $s_0$ should be such that $Y_0 \simeq 0$. 

\section*{Numerical results}
Let us first show the analysis of the dependence on $Y$ of the coefficients ${\cal C}_0$, ${\cal C}_1$, ${\cal C}_2$, where ${\cal C}_0$ is the differential cross section integrated over the tagged jets' azimuthal angles. For 
simplicity, we take the factorization and
renormalization scales equal to each other, $\mu_F=\mu_R$ (at the end of this Section we will relax this condition). We also use the PDF set MSTW2008nnlo~\cite{pdf} and the two-loop running coupling with $\alpha_s(M_Z)=0.11707.$ In order to compare with the scale dependence and values for the different coefficients obtained in previous calculations~\cite{Caporale:2012ih}, based on the same approach but without the 
collinear improvements, we select the following kinematical settings:
\begin{itemize}
\item $\sqrt{s}$=14 TeV, {\it i.e.} the LHC design value; 
\item the jet cone size has been fixed at the value $R=0.5$;
\item $|\vec k_{J_1}|=|\vec k_{J_2}|=35$ GeV.
\end{itemize}

\begin{figure}[h!]
\centering
\includegraphics[scale=0.55]{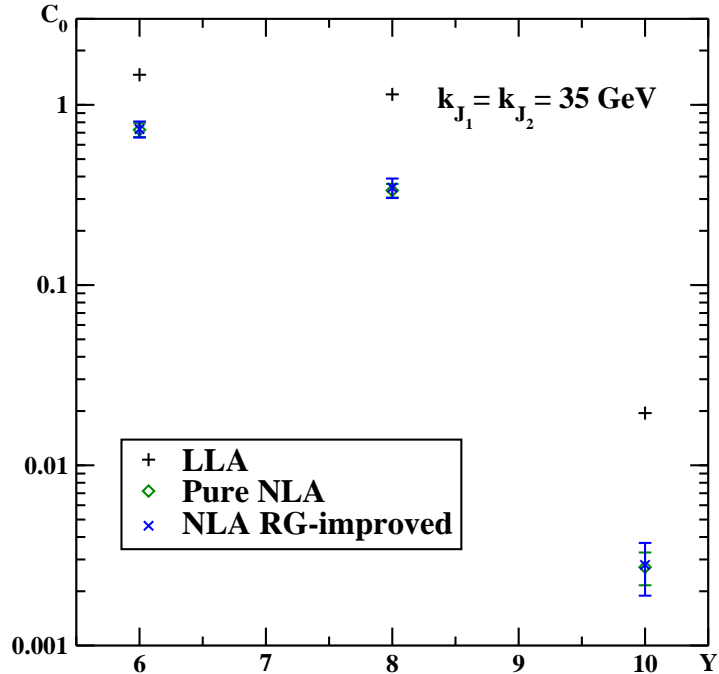}
\caption[]{$Y$ dependence of $C_0$ for $|\vec k_{J_1}|=|\vec k_{J_2}|=35$ GeV
at $\sqrt s=14$ TeV.}
\label{fig:C035}
\vspace{1cm}
\end{figure}

\begin{table}[h!]
\vspace{1cm}
\centering
\begin{tabular}{|r|c|c|c||c|c|c|}
\hline
$Y$ & $C_0^{(\rm NLA)}$ & $Y_0$ & $n_R$
    & $C_0^{(\rm RG-impr-NLA)}$ & $Y_0$ & $n_R$ \\
\hline
6  & 0.726(64)   & 1 & 2 & 0.733(75)   & 1 & 2 \\
8  & 0.335(29)   & 2 & 2 & 0.347(43)   & 2 & 2  \\
10 & 0.00272(56) & 4 & 2 & 0.00280(91) & 3 & 2  \\
\hline
\end{tabular}
\vspace{.5cm}
\caption{Values of $C_0$ in the strict NLA and with collinear improvements (RG-impr-NLA) 
for $|\vec k_{J_1}|=|\vec k_{J_2}|=35$ GeV at $\sqrt s=14$ TeV, corresponding to the data points in Fig.~\ref{fig:C035}. 
The optimal values for $Y_0$ and $n_R$
for $C_0^{(\rm NLA)}$ are given in the third and fourth columns, while
those for $C_0^{(\rm RG-impr-NLA)}$ are given in the last two columns.}
\vspace{1cm}
\label{tab:C035}
\end{table}

One immediate benefit of our collinearly-improved approach, as compared to the NLO calculation 
in~\cite{Caporale:2012ih},  is that we are able to consider also the kinematics with an asymmetric choice of the jet transverse momenta: $|\vec k_{J_1}|=20$ GeV and $|\vec k_{J_2}|=35$ GeV (we will see that there exists a region of stability in the different scales which is not present in the purely NLO approach). Following the experimental constraints described in 
Ref.~\cite{CMS}, we restrict the rapidities of the tagged jets to the region $3 \leq |y_J | \leq 5$. For our choice of forward jet rapidities, $Y$ takes values between 6 and 10. We introduce a rapidity bin size of  $\Delta y_J=0.5$ and then evaluate the sum
\begin{eqnarray}
C_n(Y) &=& \sum_j \tilde{\cal C}_n\left( (y_{J_1})_j,Y-(y_{J_1})_j\right) \, \Delta y_J
\end{eqnarray}
which runs over all the possible values of $(y_{J_1})_j$ for a given $Y$ and $\tilde{\cal C}_n (x,y)$ corresponds to the coefficient ${\cal C}_n$ where one of the jets has rapidity $x$ and the other $y$.

\begin{figure}[h!]
\vspace{1.cm}
\centering
\includegraphics[scale=0.5]{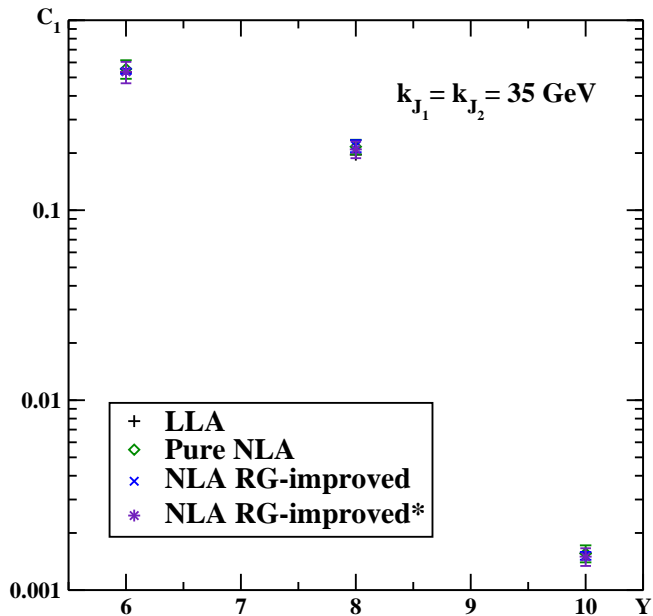}
\caption[]{$Y$ dependence of $C_1$ for $|\vec k_{J_1}|=|\vec k_{J_2}|=35$ GeV
at $\sqrt s=14$ TeV.}
\label{fig:C135}
\vspace{.5cm}
\end{figure}

\begin{table}[h!]
\vspace{1cm}
\begin{center}
\begin{tabular}{|r|c|c|c||c|c|c||c|}
\hline
$Y$ 
    & $C_1^{(\rm NLA)}$
    & $Y_0$ & $n_R$
    & $C_1^{(\rm RG-impr-NLA)}$
    & $Y_0$ & $n_R$ 
    &  $C_1^{*(\rm RG-impr-NLA)}$ \\
\hline
6 & 0.554(62) & 1 & 2 & 0.539(17)  & 0 & 1 & 0.535(69) \\
8  & 0.216(19)  & 2 & 2 & 0.218(16)  & 1 & 2 &  0.209(21)\\
10 & 0.00156(16)  & 3 & 2 & 0.001516(71) & 2 & 2 & 0.00150(16)\\
\hline
\end{tabular}
\vspace{.5cm}
\end{center}
\caption{Values of $C_1$ in the strict NLO approximation (NLA) and with collinear improvements (RG-impr-NLA) 
for $|\vec k_{J_1}|=|\vec k_{J_2}|=35$ GeV at $\sqrt s=14$ TeV, corresponding to the data points in Fig.~\ref{fig:C135}. 
The optimal values for $Y_0$ and $n_R$
for $C_1^{(\rm NLA)}$ are given in the third and fourth columns, while
those for $C_1^{(\rm RG-impr-NLA)}$ are given in the sixth and seventh columns. 
In the last column the values for $C_1$ with the collinear improvements evaluated at the optimal scales for $C_0$ are shown.}
\label{tab:C135}
\vspace{1.cm}
\end{table}

\begin{figure}[h!]
\centering
\includegraphics[scale=0.55]{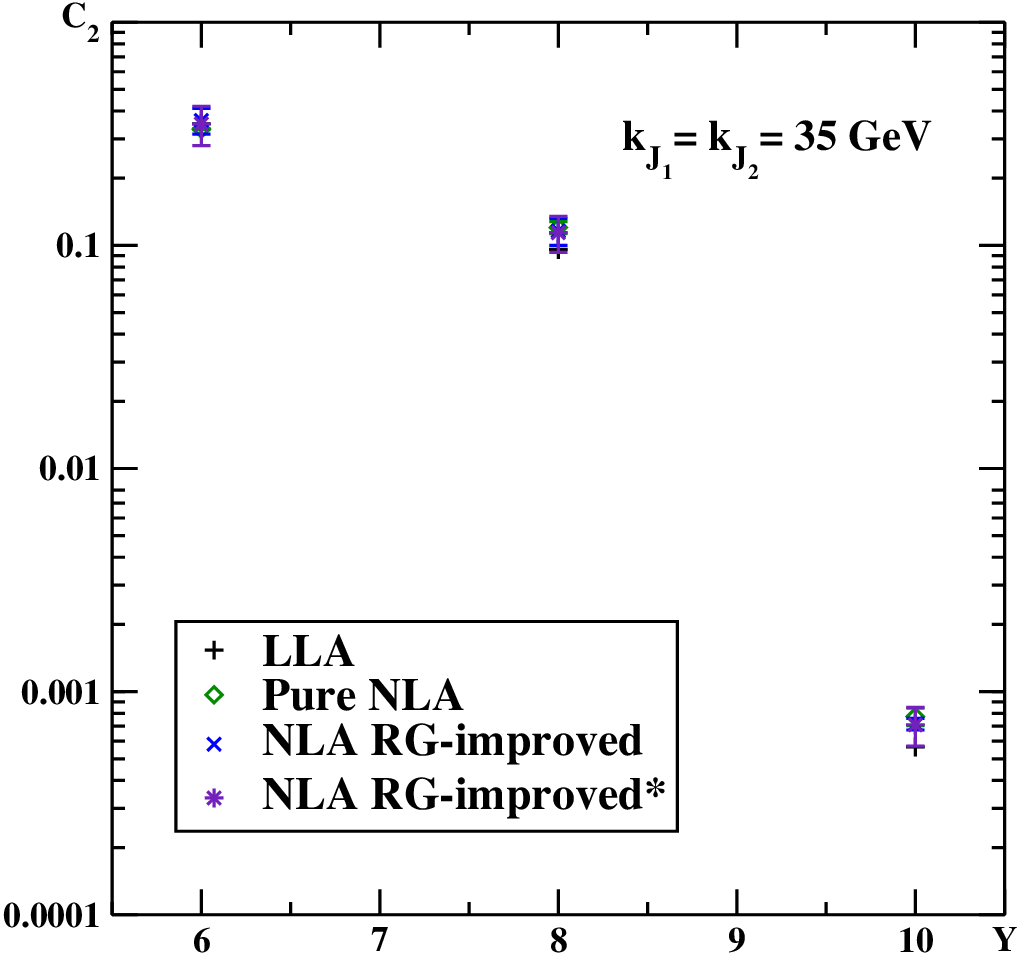}
\caption[]{$Y$ dependence of $C_2$ for $|\vec k_{J_1}|=|\vec k_{J_2}|=35$ GeV
at $\sqrt s=14$ TeV.}
\label{fig:C235}
\vspace{1cm}
\end{figure}

\begin{table}[h!]
\vspace{1cm}
\begin{center}
\begin{tabular}{|r|c|c|c||c|c|c||c|}
\hline
$Y$ 
    & $C_2^{(\rm NLA)}$
    & $Y_0$ & $n_R$
    & $C_2^{(\rm RG-impr-NLA)}$
    & $Y_0$ & $n_R$ 
    &  $C_2^{*(\rm RG-impr-NLA)}$ \\
\hline
6 & 0.3320(18) & 0 & 1.5 & 0.326(15)  & 0 & 1 & 0.350(70) \\
8  & 0.1203(74)  & 2 & 2.5 &  0.116(16)  & 2 & 3 &  0.114(21)\\
10 & 0.000774(69)  & 4 & 4 & 0.000716(43) & 2 & 2 & 0.00071(14)\\
\hline
\end{tabular}
\vspace{.5cm}
\end{center}
\caption{
Values of $C_2$ in the strict NLO approximation (NLA) and with collinear improvements (RG-impr-NLA) 
for $|\vec k_{J_1}|=|\vec k_{J_2}|=35$ GeV at $\sqrt s=14$ TeV, corresponding to the data points in Fig.~\ref{fig:C235}. 
The optimal values for $Y_0$ and $n_R$
for $C_2^{(\rm NLA)}$ are given in the third and fourth columns, while
those for $C_2^{(\rm RG-impr-NLA)}$ are given in the sixth and seventh columns. 
In the last column the values for $C_2$ with the collinear improvements evaluated at the optimal scales for $C_0$ are shown.}
\label{tab:C235}
\vspace{1cm}
\end{table}

\begin{figure}[h!]
\centering
\includegraphics[scale=0.55]{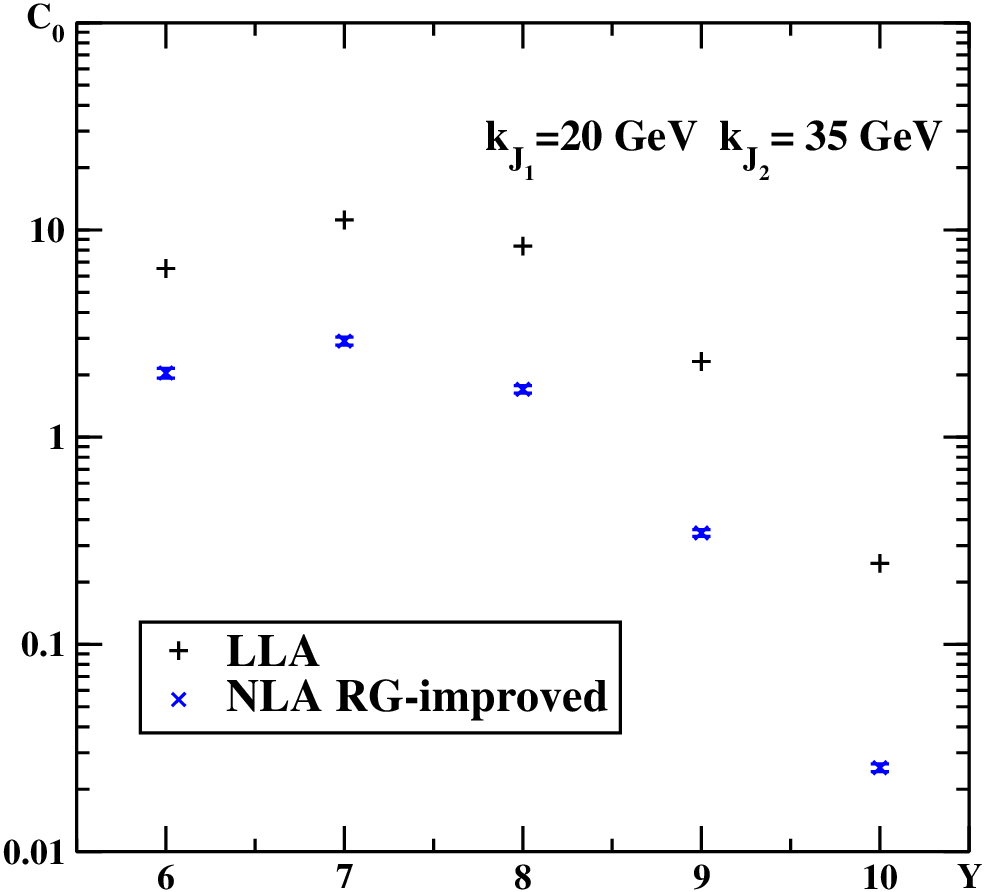}
\caption[]{$Y$ dependence of $C_0$ for $|\vec k_{J_1}|=20$ GeV, $|\vec k_{J_2}|=35$ GeV
at $\sqrt s=14$ TeV.}
\label{fig:C02035}
\vspace{1cm}
\end{figure}

\begin{table}[h!]
\vspace{1cm}
\begin{center}
\begin{tabular}{|r|c|c|c|}
\hline
$Y$ 
    & $C_0^{(\rm RG-impr-NLA)}$ & $Y_0$ & $n_R$ \\
\hline
6  & 2.04(11)   & 2 & 3 \\
7  & 2.91(13)    & 1 & 2.5 \\ 
8  & 1.703(70)   & 2 & 2.5  \\
9  & 0.345(13)  &  1.5 & 3 \\ 
10 & 0.0254(11) & 2.5 & 3  \\
\hline
\end{tabular}
\vspace{.5cm}
\end{center}
\caption{Values of $C_0$ at NLO with collinear improvements for $|\vec k_{J_1}|=20$ GeV and $|\vec k_{J_2}|=35$ GeV at $\sqrt s=14$ TeV, corresponding to the data points in Fig.~\ref{fig:C02035}. The
optimal values of $Y_0$ and $n_R$ are given in the last two columns.}
\label{tab:C02035}
\vspace{1cm}
\end{table}

\begin{figure}[h!]
\centering
\includegraphics[scale=0.55]{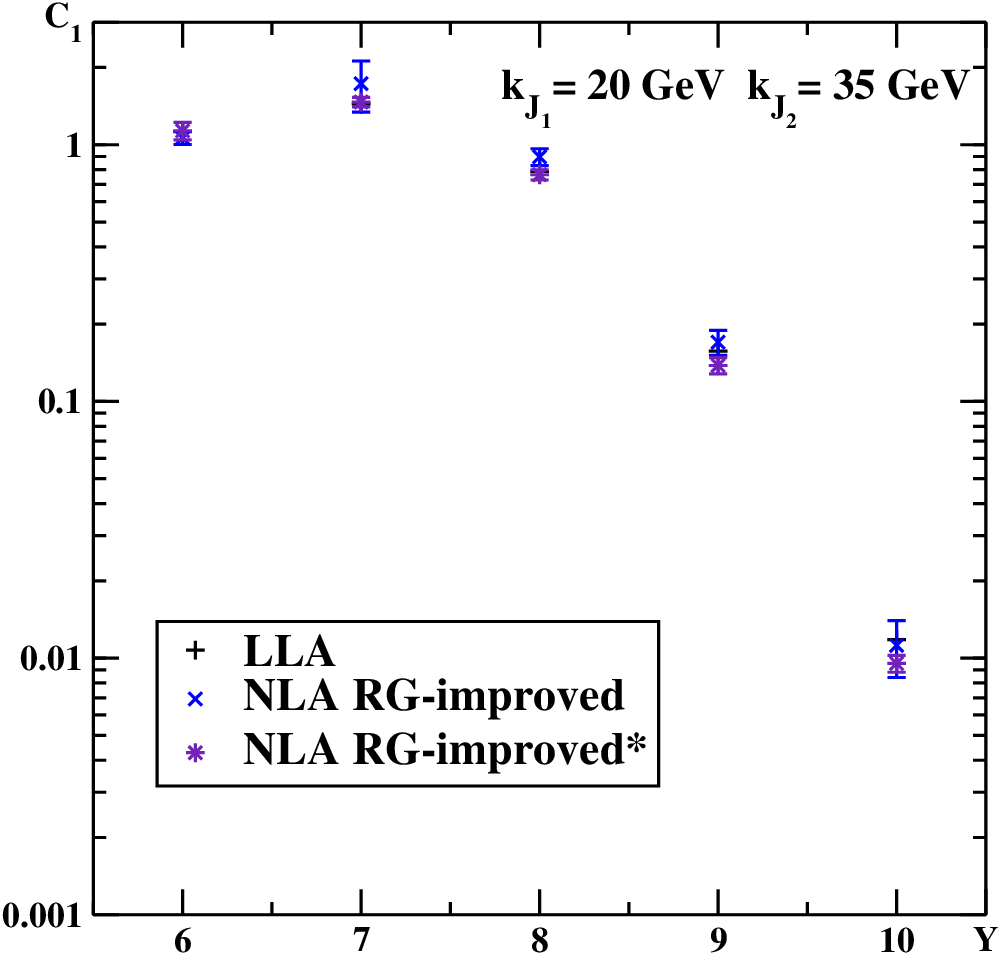}
\caption[]{$Y$ dependence of $C_1$ for $|\vec k_{J_1}|=20$ GeV, $|\vec k_{J_2}|=35$ GeV
at $\sqrt s=14$ TeV.}
\label{fig:C12035}
\vspace{1cm}
\end{figure}

\begin{table}[h!]
\vspace{1cm}
\begin{center}
\begin{tabular}{|r|c|c|c||c|}
\hline
$Y$  
   & $C_1^{(\rm RG-impr-NLA)}$
    & $Y_0$ & $n_R$ 
    &  $C_1^{*(\rm RG-impr-NLA)}$ \\
\hline
6 & 1.384(88)  & 1 & 1 & 1.133(89) \\
7 & 1.73(39) & 1 & 1 & 1.466(63)  \\
8  & 0.897(68)  & 1 & 1 &  0.764(35)\\
9  & 0.170(19) & 2 & 1 &  0.138(10) \\
10 & 0.0112(28) & 3 & 1 & 0.00953(72)\\
\hline
\end{tabular}
\vspace{.5cm}
\end{center}
\caption{Values of $C_1$ in the NLA
and in the NLA with
collinear improvement for $|\vec k_{J_1}|=20$ GeV and $|\vec k_{J_2}|=35$
GeV at $\sqrt s=14$ TeV, corresponding to the data points in
Fig.~\ref{fig:C12035}. The optimal values of $Y_0$ and
$n_R$ for $C_1^{(\rm RG-impr-NLA)}$ are
given in the third and fourth columns. In the last column there are the values obtained for $C_1$ with the collinear improvement for the same of optimal scales of $C_0$.}
\label{tab:C12035}
\vspace{1cm}
\end{table}

\begin{figure}[h!]
\centering
\includegraphics[scale=0.55]{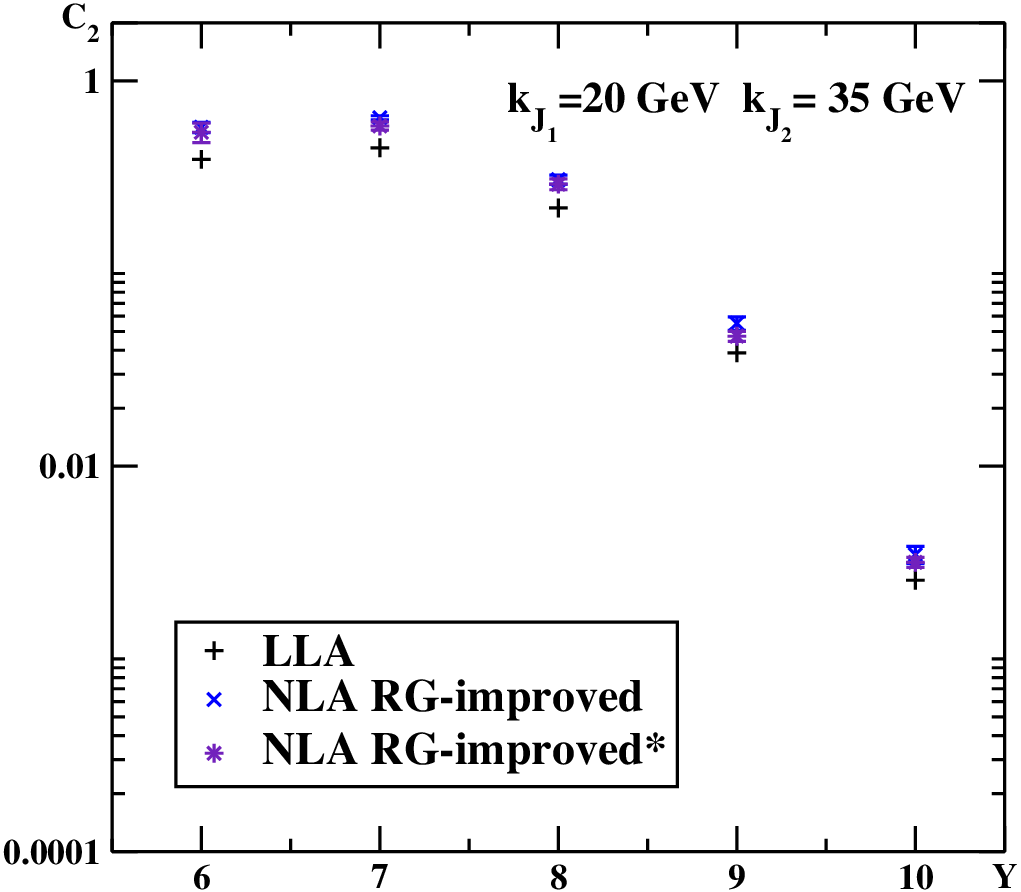}
\caption[]{$Y$ dependence of $C_2$ for $|\vec k_{J_1}|=20$ GeV, $|\vec k_{J_2}|=35$ GeV
at $\sqrt s=14$ TeV.}
\label{fig:C22035}
\vspace{1cm}
\end{figure}

\begin{table}[h!]
\vspace{1cm}
\begin{center}
\begin{tabular}{|r|c|c|c||c|}
\hline
$Y$  
   & $C_2^{(\rm RG-impr-NLA)}$
    & $Y_0$ & $n_R$ 
    &  $C_2^{*(\rm RG-impr-NLA)}$ \\
\hline
6 & 0.574(35)  & 1 & 1 & 0.541(63) \\
7 & 0.643(16) & 1 & 0.75 & 0.583(28)  \\
8  & 0.307(17)  & 1 & 1 &  0.291(19)\\
9  & 0.0552(44) & 2 & 1 &  0.0473(28) \\
10 & 0.00348(36) & 2 & 1 & 0.00317(19) \\
\hline
\end{tabular}
\vspace{.5cm}
\end{center}
\caption{Values of $C_2$ in the NLA
and in the NLA with
collinear improvement for $|\vec k_{J_1}|=20$ GeV and $|\vec k_{J_2}|=35$
GeV at $\sqrt s=14$ TeV, corresponding to the data points in
Fig.~\ref{fig:C22035}. The optimal values of $Y_0$ and
$n_R$ for
 $C_2^{(\rm NLA /RG-impr)}$ are
given in the third and fourth columns. In the last column there are the values obtained for $C_2$ with the collinear improvement for the same of optimal scales of $C_0$.}
\label{tab:C22035}
\vspace{1cm}
\end{table}

Our expressions for the coefficients $C_n$, when expanded at NLO (${\cal O} (\alpha_s^2)$), 
do not have any dependence on the renormalization, 
$\mu_R$, and energy, $s_0$, scales (as we have already indicated, we have chosen the factorization scale to equal $\mu_R$). However, when exponentiating 
the BFKL kernel, following bootstrap, higher order terms, beyond NLO, are introduced and generate a residual dependence on these scales. This dependence would cancel again order by order in perturbation theory if we had the BFKL kernel and jet vertices calculated at higher orders. In a purely NLO approach 
(with the conformal invariant parts of the kernel exponentiated) the dependence on these scales is larger than when introducing the collinear 
improvements. This is what we will show with our numerical results, where we will see that the regions of stationary values in the multidimensional scale space is closer to the physical scales in the problem (the jets' $p_t^2$) in the latter case than in the former. Following previous works~\cite{mesons, Caporale:2011cc,Caporale:2012ih}, in our analysis we will use an adaptation of the
{\it principle of minimal sensitivity} (PMS)~\cite{Stevenson}, where we consider as optimal choices for $\mu_R$ and $s_0$ those values for which the physical quantity under examination exhibits the minimal sensitivity to changes in both of these scales.
Without using the RG-improved kernel the optimal choices for these parameters, when Y grows, turned out to be quite far from the kinematical scales of the process \cite{Caporale:2012ih}. Let us see how the inclusion of the collinear improvement leads to more ``natural" values for the optimal scales (similar results were found in Ref.~\cite{mesons} in the context of light vector meson production).

In our search for optimal values, we took integer values for $Y_0$ while for $\mu_R$ we look for integer multiples of
$\sqrt{|\vec k_{J_1}| |\vec k_{J_2}|}$ in the form
\beq{}
\mu_R=n_R \sqrt{|\vec k_{J_1}| |\vec k_{J_2}|}\,.
\eeq
In this way, the systematic uncertainty of the optimization procedure stems from the resolution of a grid in the  $Y_0$ -- $n_R$  plane and we consider as ``natural" values of $n_R$ those close to one.

Let us first discuss the results for the symmetric kinematics. Filling a grid in the  $Y_0$ -- $n_R$  plane we found that a stationary point 
could always be singled out.
Our results, in
$\left[\rm{nb} / \rm{GeV}^2\right]$ units, are presented in
Figs.~\ref{fig:C035}--\ref{fig:C135}--\ref{fig:C235} and in
Tables~\ref{tab:C035}--\ref{tab:C135}--\ref{tab:C235}. 
For the coefficients we find the optimal values using the PMS and present the values of the optimal scales obtained  (last columns of Tables~\ref{tab:C135} and \ref{tab:C235}, corresponding to the results labeled by ``RG-improved*" in Figs.~\ref{fig:C135}--\ref{fig:C235}).
We can see that for the optimal scales there is a small shift towards naturalness, in particular for high values of $Y$. Even if this effect is less evident than in~\cite{mesons}, it shows that the collinear improvements stabilize the perturbative series.
Nevertheless, it is interesting to note that the actual values of the coefficients are in good agreement with the canonical NLO results (they overlap within the error bars), even if the RG-improved results are a bit higher for $C_0$ and lower for $C_{n>0}$, as it is expected since the RG improvements make the asymptotic Pomeron intercept ($n=0$) to be larger without modifying the $n>0$ intercepts. This is different to what we found in the 
case of the electroproduction of light vector mesons in~\cite{mesons}, where both approaches generated very different 
results at the observable level. We believe the main reason for this is that in the case of Mueller-Navelet jets the actual phase space 
for multijet production is highly constrained by the PDFs, which prevent our cross sections from growing 
at asymptotic values of $Y$. It is also noteworthy that the values for $C_1$ and $C_2$ obtained with the PMS overlap with the values obtained when being evaluated at the ``optimal" scales found for $C_0$.

For the asymmetric case, with $|\vec k_{J_1}|=20$ GeV and $|\vec k_{J_2}|=35$ GeV, we present our results
in Figs.~\ref{fig:C02035}--\ref{fig:C12035}--\ref{fig:C22035}  and
Tables~\ref{tab:C02035}--\ref{tab:C12035}--\ref{tab:C22035}. For $C_1$ and $C_2$ we again find the optimal values using the PMS and we also show the values corresponding to the optimal scales obtained for $C_0$ (last column of Tables ~\ref{tab:C12035} and \ref{tab:C22035}, together with the results called ``RG-improved*" in Figs.~\ref{fig:C12035}--\ref{fig:C22035} ).
Let us remark that in the NLO approach it was not possible to find a stability region in the $n_R$ -- $Y_0$ plane 
and that, importantly, the inclusion of the RG-improved kernel proved to be very useful (a similar situation happened for the vector meson production case~\cite{mesons}). In our search  for optimal scale values for $C_0$ and $C_1$ we always found a stability region, while for the coefficient $C_2$ only for a few values of $Y$ we could find a
stationary point. In other cases we found a local maximum only in the
direction of one of the two parameters. When this happened, we took as
``optimal'' value for the observable the one exhibiting the least standard deviation from the values taken in the nearest neighboring points in the chosen grid.
We can see that the ``optimal" values for the parameters are quite ``natural", in particular for $C_1$ and $C_2$.
On the other hand, the obtained PMS values for $C_1$ and $C_2$ and those corresponding to the ``optimal" scales for $C_0$ 
differ from each other more than in the case of a symmetric kinematics, but still overlapping within the error bars.

\begin{figure}[tb]
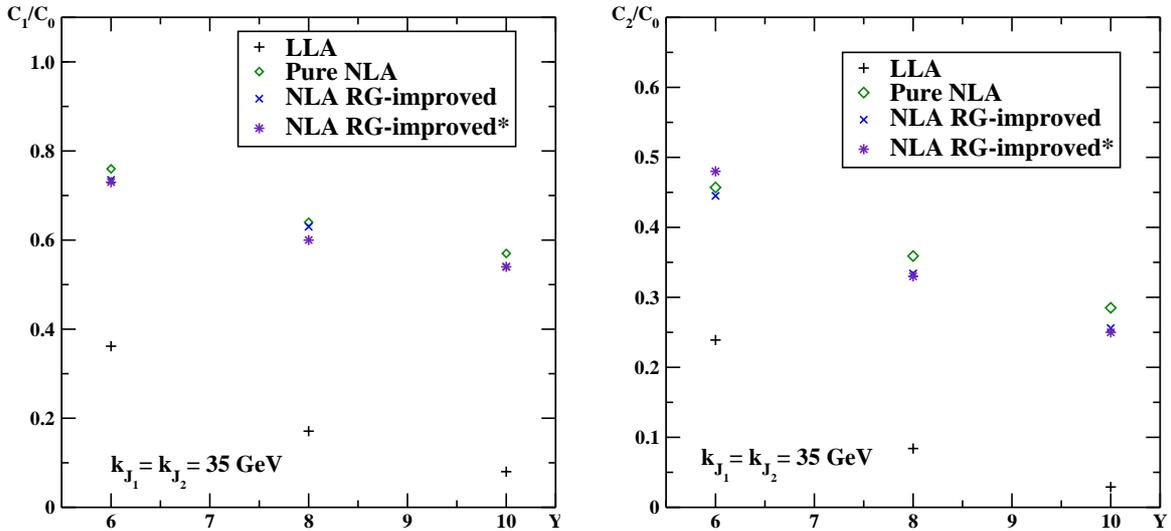

\hspace{-.9cm}
\includegraphics[scale=0.43]{C1C035_SE.eps} \hspace{0.5cm}\includegraphics[scale=0.43]{C2C035_SE.eps}
\caption[]{$Y$ dependence of $C_1/C_0$ (left)
and $C_2/C_0$ (right) for $|\vec k_{J_1}|=|\vec k_{J_2}|=35$ GeV at
$\sqrt s=14$ TeV. }
\label{fig:C12C035_SE}
\end{figure}

Having the complete information about the coefficients $C_m$ we now present the analysis of the $Y$ dependence of the moments of the azimuthal decorrelation, which read
\begin{eqnarray}
\langle \cos(m \phi) \rangle &=& \frac{{\cal C}_m}{{\cal C}_0}.
\label{azcorr}
\end{eqnarray}

We start by presenting the results for the symmetric kinematics. Filling a grid in the  $n_R$ -- $Y_0$ plane we found that a stationary point could always be singled out. Our results are shown in different figures. In Fig.~\ref{fig:C12C035_SE} we present $\langle \cos(\phi) \rangle = C_1/C_0$ and  $\langle \cos( 2 \phi) \rangle  = C_2/C_0$ as a function of $Y$. We observe a strong decorrelation as $Y$ increases, generated by the wealth of radiation produced by the iteration of the BFKL kernel. This decorrelation is largely 
 reduced, with respect to the LO calculation, when the NLO corrections are introduced, indicating that the amount of real emission is much smaller in this approximation. It is interesting to note that introducing collinear 
 improvements in the NLO result does not have a very big effect, slightly reducing the amount of azimuthal-angle correlation 
 between the two tagged jets. This is natural since the collinear regions of phase space are tamed by having 
 two transverse momenta of the same magnitude ($|\vec k_{J_1}|=|\vec k_{J_2}|=35$ GeV). As we explore more asymmetric configurations the impact of the collinear resummation is larger, allowing for stability regions not found in the 
 pure NLO case. 
 
 We have calculated both ratios $C_{m>0}^{(\rm RG-impr-NLA)}/C_0^{(\rm RG-impr-NLA)}$ and $C_{m>0}^{*(\rm RG-impr-NLA)}/C_0^{(\rm RG-impr-NLA)}$ (the results called ``RG-improved*" in Fig.~\ref{fig:C12C035_SE}), with the latter generating a slightly lower correlation at larger rapidities. 
 
\begin{figure}[tb]
\centering
\includegraphics[scale=0.55]{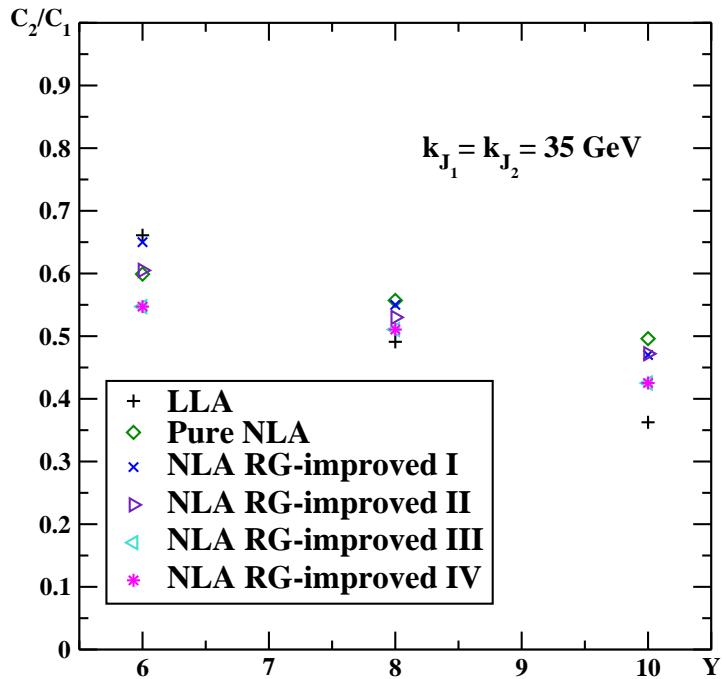}
\caption[]{$Y$ dependence of $C_2/C_1$ for $|\vec k_{J_1}|=|\vec k_{J_2}|=35$ GeV (left).
at $\sqrt s=14$ TeV.}
\label{fig:C2C135aver}
\end{figure}
It is noteworthy to indicate that the origin of the large difference between the LO and NLO results (also with all-order improvements) is due to the influence of $C_0$ on these observables. The reason for this, as we have already mentioned, is that $C_0$ does not enjoy a good perturbative convergence because it is related to the 
 conformal spin $n=0$. However, as the $C_{n>0}$ do have good perturbative convergence within the BFKL formalism, the following ratios were proposed as the {\bf ideal BFKL observables} 
 in~\cite{Vera:2006un,Vera:2007kn}:
\begin{eqnarray}
{\cal R}_{mn} &\equiv& \frac{\langle \cos{(m \Delta \phi)} \rangle}{\langle \cos{(n \Delta \phi)}\rangle} 
~=~ \frac{C_m}{C_n}.
\end{eqnarray}
They are free from $n=0$ contributions if $m,n \neq 0$. Let us now show (for ${\cal R}_{21}$,  but the qualitative behaviour is similar for other ratios) that the theoretical prediction is very similar at LO, NLO and with collinear improvements, making of ${\cal R}_{mn}$ a robust test of the whole BFKL formalism. This is shown, for  a symmetric configuration, in Fig.~\ref{fig:C2C135aver}.  In order to gauge the theoretical uncertainty of our results we performed four different calculations:
\begin{itemize}
\item $C_2^{(\rm RG-impr-NLA)}/C_1^{(\rm RG-impr-NLA)}$ (``RG-improved I" in Fig.~\ref{fig:C2C135aver}).
\item $C_2^{*(\rm RG-impr-NLA)}/C_1^{*(\rm RG-impr-NLA)}$ (``RG-improved II" in Fig.~\ref{fig:C2C135aver}).
\item ${(C_2/C_1)}^{(\rm \mu_F \, fixed)}$, where we have relaxed the condition $\mu_F = \mu_R$ and fixed 
$\mu_F=|\vec k_{J_1}|$ for one of the hadrons and $\mu_F= | \vec k_{J_2}|$ for the other, using the PMS to find the best values for $\mu_R$ and $Y_0$ (denoted by ``RG-improved III" in Fig.~\ref{fig:C2C135aver}). The values can be found in Table~\ref{tab:C2C12035} (left).
\item $(C_2/C_1)^{(\rm \mu_F=\mu_R)}$, ``RG-improved IV" in Fig.~\ref{fig:C2C135aver}, where we restate the condition $\mu_F = \mu_R$ using the same optimal scales as for $(C_2/C_1)^{(\rm \mu_F \, fixed)}$, without finding any deviation in the value of the observable (in this case we could not find any reasonable stability region with optimal scales and this is why we chose the same ones as in the previous point). 
\end{itemize}

\begin{table}[tb]
\begin{center}
\scalebox{1}{
\begin{tabular}{|r|c|c|c|}
\hline
$Y$ 
    & ${\cal R}_{21{\;\rm sym}}$
    & $Y_0$ & $n_R$
    \\
\hline
6 & 0.5471 & 1.5 & 1  \\
8  & 0.5105  & 1.5 & 1 \\
10 & 0.4253  & 0 & 1 \\
\hline
\end{tabular}
\hspace{.5cm}
\begin{tabular}{|r|c|c|c|c|}
\hline
$Y$ 
    & ${\cal R}_{21{\;\rm asym}}^{(\rm \mu_F \, fixed)}$
    & ${\cal R}_{21{\;\rm asym}}^{(\rm \mu_F=\mu_R)}$
    & $Y_0$ & $n_R$  \\
\hline
6 & 0.3954 & 0.3940 & 0 & 1   \\
7 & 0.3567 & 0.3548 & 0 & 1  \\
8 &  0.3258  & 0.3267  & 0 & 1 \\
9  & 0.2860 & 0.2992  & 0 & 1 \\
10 & 0.2831 & 0.2848 & 0 & 1 \\
\hline
\end{tabular}
}
\end{center}
\caption{Right table: values for $C_2/C_1$ -corresponding to fig.~\ref{fig:C2C12035} (right)- using collinearly improved NLL resummation with asymmetric configuration $|\vec k_{J_1}|=20$ GeV and $|\vec k_{J_2}|=35$ GeV
at $\sqrt s=14$ TeV, setting $\mu_{F_1}=|\vec k_{J_1}|$ and $\mu_{F_2}=|\vec k_{J_2}|$ (second column) and fixing $\mu_{F_1}=\mu_{F_2}=\mu_R$ (third column).  
The optimal values of $Y_0$ and
$n_R$, given in the last two columns, are the same for the two cases. Left table: same for symmetric configuration $|\vec k_{J_1}|=|\vec k_{J_2}|=35$ GeV, with ${\cal R}_{21{\;\rm sym}}^{(\rm \mu_F \, fixed)}={\cal R}_{21{\;\rm sym}}^{(\rm \mu_F=\mu_R)}$.}
\label{tab:C2C12035}
\end{table}


\begin{figure}[t]
\vspace{-1cm}
\centering
\includegraphics[scale=0.55]{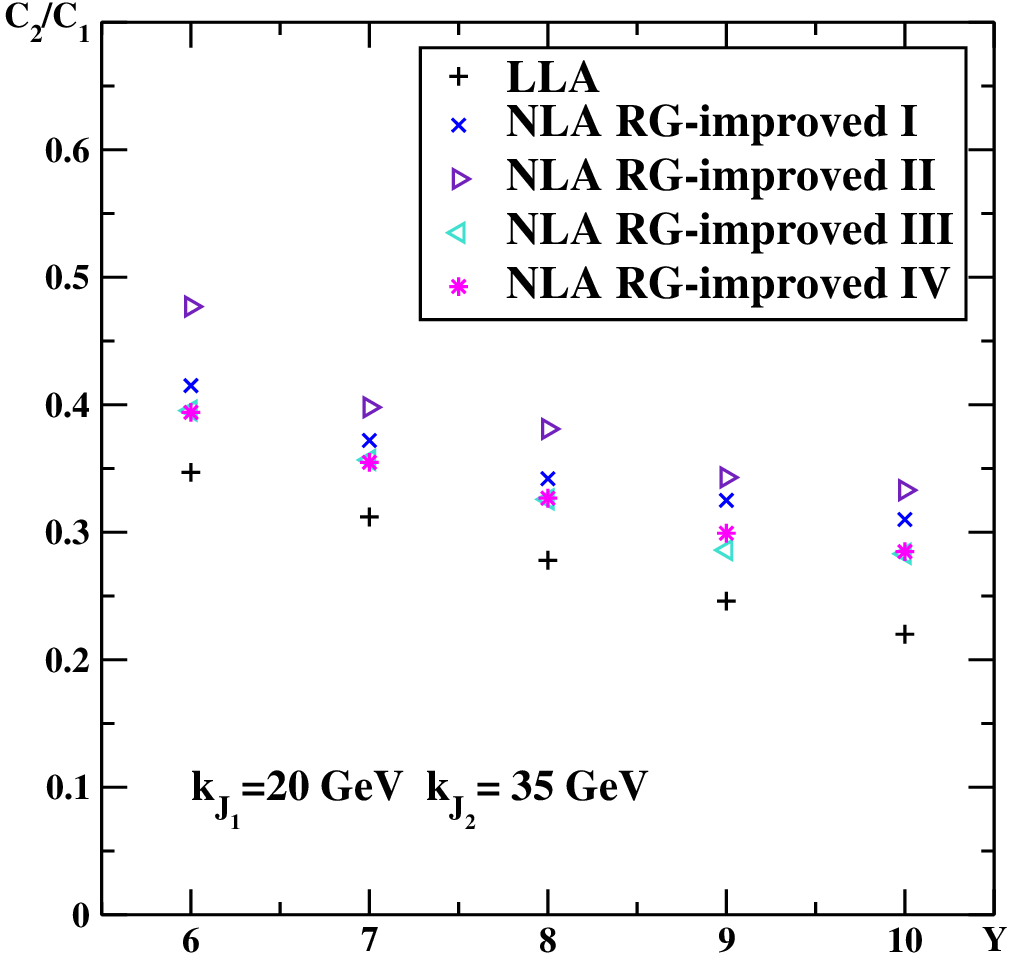}
\caption[]{$Y$ dependence of $C_2/C_1$ for $|\vec k_{J_1}|=20$ GeV, $|\vec k_{J_2}|=35$ GeV
at $\sqrt s=14$ TeV.}
\label{fig:C2C12035}
\vspace{1cm}
\end{figure}
\begin{figure}[h!]
\hspace{-.9cm}\includegraphics[scale=0.43]{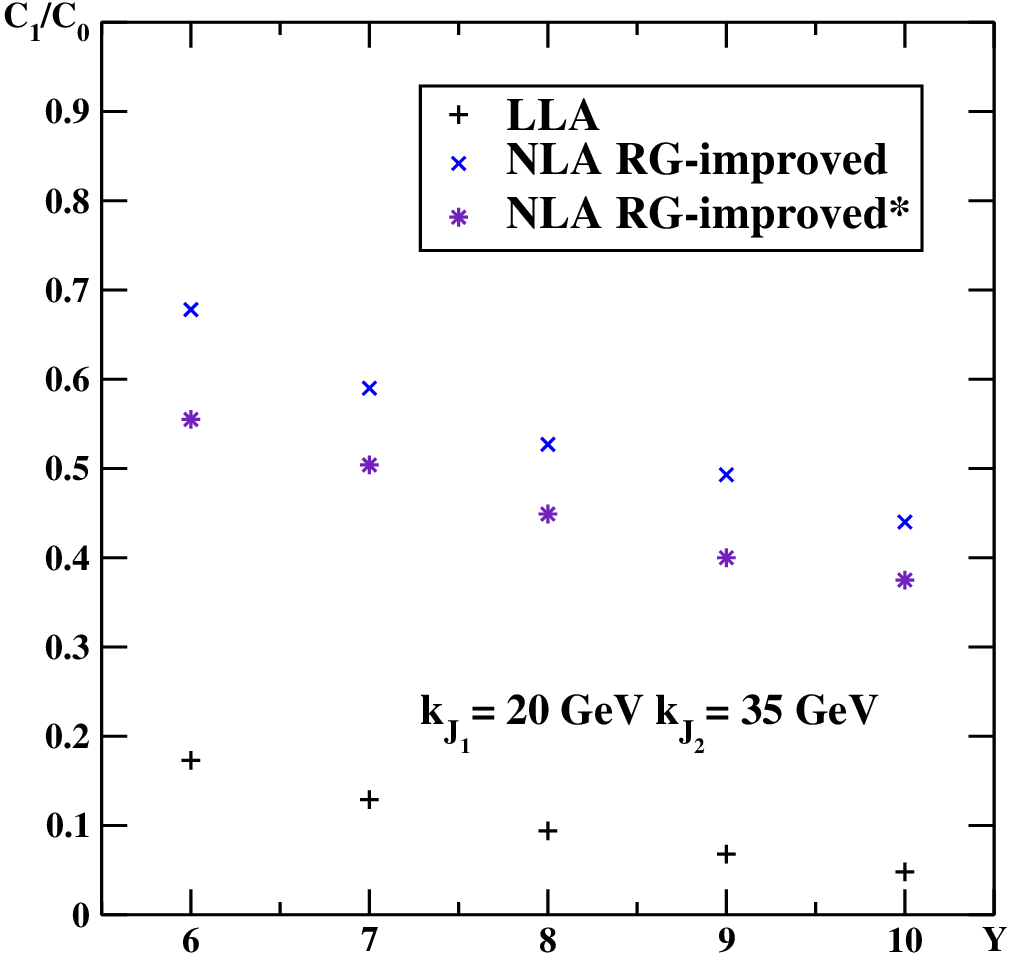} \hspace{.5cm}\includegraphics[scale=0.43]{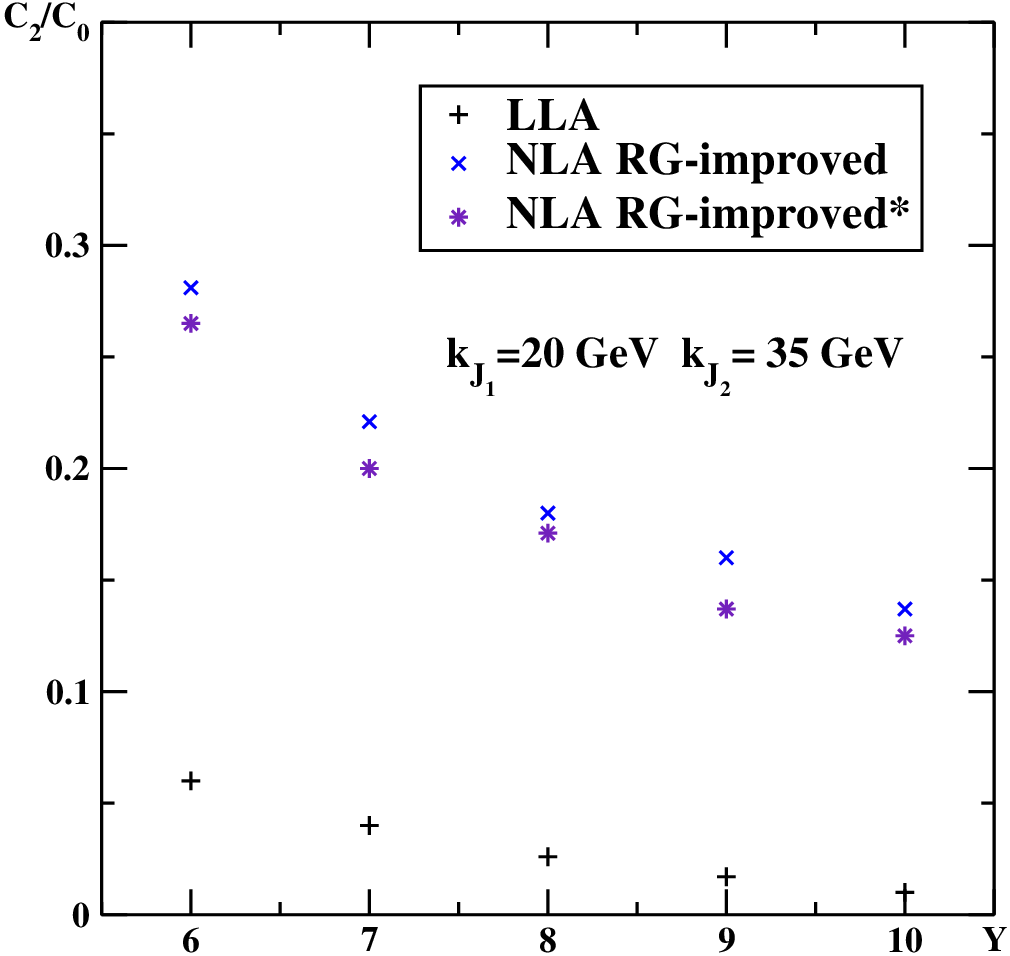}
\caption[]{$Y$ dependence of $C_1/C_0$ (left)
and $C_2/C_0$ (right) for $|\vec k_{J_1}|=20$ GeV, $|\vec k_{J_2}|=35$ GeV at
$\sqrt s=14$ TeV. }
\label{fig:C1C02035}
\end{figure}

Let us conclude our analysis with the asymmetric case, with $|\vec k_{J_1}|=20$ GeV and $|\vec k_{J_2}|=35$ GeV, where the collinear effects are more noticeable. The labels and analysis are as for the symmetric kinematics and the results are shown in Fig.~\ref{fig:C1C02035} for $C_{1,2} / C_0$. The ratio ${\cal R}_{21}$ is presented in Fig.~\ref{fig:C2C12035} together with the corresponding values of the optimal scales we could find in Table~\ref{tab:C2C12035}. We found 
the same lack of stable regions when setting $\mu_F = \mu_R $ which we have solved by relaxing this condition and taking $\mu_F = |\vec k_{J_1}|$ and $\mu_F = |\vec k_{J_2}|$ as the 
factorization scales associated to each of the hadrons. This is a very fortunate choice since it 
creates a stability region at the ``very natural" point  $(Y_0, n_R)=(0,1)$ which is invariant under 
changes in $Y$. 


\newpage

\section*{Conclusions}

The calculation and numerical implementation of the NLO forward jet vertices, together with the NLO gluon Green function, 
offers the opportunity to investigate in detail the perturbative convergence of the BFKL program in the hadroproduction of Mueller-Navelet jets. This happens at a time when we have a wealth of experimental data produced at the Large Hadron Collider. Due to the theoretically sound bootstrap property of QCD 
at high energies it is possible to go beyond the standard field theory calculations of scattering amplitudes and use effective ``reggeized" degrees of freedom to make predictions in the high energy limit. This approach generates some dependence on renormalization, factorization and energy scales, to all orders, which we can minimize looking for regions of maximal stability in the 
variation of these parameters. If exact higher order corrections were calculated, the values at these regions for our observables would be good candidates where the scale independent values would finally lie. A good hint that we have reliable predictions comes from the 
fact that the regions of stability for our scales are not far from the ``natural" values (the typical squared transverse momentum of the tagged jets). Since the BFKL expansion needs to be stabilized in the collinear regions, beyond the original quasi-multi-Regge kinematics where the original approximations when calculating the amplitudes lie, it is natural to expect that the ``optimal" values of the free scales in our calculations will be more ``natural" when the BFKL kernel is collinearly improved with an all-order resummation designed to properly cover a larger region of phase space. This is what we have shown in this work. 
The effect of the collinear improvements is not as dramatic as in other fully NLO calculations (electroproduction of light vector mesons) since in Mueller-Navelet jets the parton distribution functions play a very strong role when the rapidity separation of the two jets is large. In this way the actual values of the observables (cross sections, azimuthal angle decorrelations and ratios of them) 
do not vary much when using a strict NLO approach or a collinearly-improved one. This is particularly true when targetting configurations with tagged jets of similar squared transverse momentum, but not so much for asymmetric configurations, where the 
collinear improvements are actually needed to obtain stability regions at all (in a pure NLO analysis this was not possible). 

As a future line of research we find it interesting to extend our investigations to find stability regions in the multiparameter scale 
space using Monte Carlo event generators implementing the NLO BFKL 
dynamics~\cite{Chachamis:2011rw} directly in transverse momentum space. 
This will allow us to gauge in detail how different treatments of the running of the coupling might affect the choice of ``optimal" scales and how far into softer regions~\cite{Hentschinski:2012kr} we can push our calculations. 
\bigskip \\
\noindent {\bf Acknowledgements}
\medskip\\
F.C. thanks the Instituto de F\'isica Te\'orica UAM/CSIC for the warm hospitality. We acknowledge partial support from the European Comission under contract LHCPhenoNet (PITN-GA-2010-264564), the Comunidad de Madrid through HEPHACOS S2009/ESP-1473, and MICINN (FPA2010-17747) and Spanish MINECOs Centro de Excelencia Severo Ochoa Programme under grant SEV- 2012-0249. The work of F.C. was supported by European Commission, European Social Fund and Calabria Region, that disclaim any liability for the use that can be done of the information provided in this paper.

\end{document}